\documentclass{PoS}

\usepackage{amssymb,fontenc,mathptmx,graphicx,amsmath}

\title{Future limits on isotropic Lorentz violation in the photon sector
       from UHECRs and TeV gamma rays}

\ShortTitle{Future limits on isotropic LV from UHECRs and TeV $\gamma$ rays}

\author{\speaker{F.R. Klinkhamer}\\
        Institute for Theoretical Physics, University of Karlsruhe,
        Karlsruhe Institute of Technology, 76128 Karlsruhe, Germany\\
        E-mail: \email{frans.klinkhamer@kit.edu}}

\abstract{Present and future ultra-high-energy-cosmic-ray facilities
(e.g., the Pierre Auger Observatory with South and North components)
and TeV-gamma-ray telescope arrays (e.g., HESS or VERITAS and CTA)
have the potential to set stringent indirect bounds on the nine
Lorentz-violating parameters of nonbirefringent modified Maxwell
theory minimally coupled to standard Dirac theory.
Theoretically, the most interesting case is isotropic Lorentz
violation, which is described by a single parameter [taken to
vanish for the standard Lorentz-invariant theory].
It appears possible to obtain in the future an upper (lower)
indirect bound on this single isotropic Lorentz-violating
parameter at the $+10^{-21}$ $\big(-10^{-17\,}\big)$ level.
Comparison is made with existing and future direct bounds
from laboratory experiments. The possible physics implications of
upper bounds at the $10^{-21}$ level are briefly discussed.}

\FullConference{25th Texas Symposium on Relativistic Astrophysics \\
                 December 6--10, 2010\\
                 Heidelberg, Germany}

\begin{document}

\section{Isotropic Lorentz violation in the photon sector}
\label{sec:Isotropic-LV}

If Lorentz violation (LV) occurs somewhere in the theory of
elementary particles and their interactions, then it can be expected
to feed also into the photon sector.
This makes the search of possible Lorentz-violating effects
in the photon sector of substantial interest, especially as photons
can be measured accurately and in a variety of physical systems.

Consider the isotropic modified Maxwell
theory~\cite{ChadhaNielsen1983,KosteleckyMewes2002}
minimally coupled to the standard Dirac theory of
a spin--$\textstyle{\frac{1}{2}}$ particle with charge $e$ and mass $M$.
The Lagrange density of this particular modification
(``deformation'') of quantum electrodynamics (QED) is given by
\begin{subequations}\label{eq:L-modQED-modMax}
\begin{eqnarray}\label{eq:L-modQED}
\mathcal{L}_\text{\,modQED}[c,\widetilde{\kappa}_\text{tr},M,e]&=&
\mathcal{L}_\text{\,modMaxwell}[c,\widetilde{\kappa}_\text{tr}] +
\mathcal{L}_\text{\,Dirac}[c,M,e]\,,\\[2mm]
\mathcal{L}_\text{\,modMaxwell}[c,\widetilde{\kappa}_\text{tr}](x) &=&
\textstyle{\frac{1}{2}}\,\Big( (1+\widetilde{\kappa}_\text{tr})\,|\mathbf{E}(x)|^2 -
(1-\widetilde{\kappa}_\text{tr})\,|\mathbf{B}(x)|^2 \Big)\,,
\label{eq:L-modMax}
\end{eqnarray}
\end{subequations}
with Cartesian spacetime coordinates $(x^\mu)=(c\,t,\,x^1,\,x^2,\,x^3)$
and the standard Lagrange density of a Dirac particle from the
textbooks, some of which are listed in Ref.~\cite{KlinkhamerSchreck2010}.

This theory is gauge-invariant, CPT--even, and power-counting
renormalizable, but, for $\widetilde{\kappa}_\text{tr}\ne 0$,
violates the Lorentz boost invariance while maintaining rotational
invariance in a preferred reference frame.
One possible reference frame is the one
with an isotropic Cosmic Microwave Background. Here, though,
the usual choice of the experimentalists is followed by
employing the sun-centered celestial equatorial frame.

Two questions immediately arise. First, is the modified-QED theory
theoretically consistent for all values of the parameter
$\widetilde{\kappa}_\text{tr}$ or is there a
restricted parameter domain?
Second, the modified-QED theory \eqref{eq:L-modQED-modMax} is
formulated in a flat spacetime, but how about gravity?
Very briefly, the answers are as follows.
The theory \eqref{eq:L-modQED-modMax} is
consistent (i.e., has microcausality and unitarity) only for
parameters in a restricted domain~\cite{KlinkhamerSchreck2010},
\begin{equation}\label{eq:domain}
\widetilde{\kappa}_\text{tr} \in (-1,\,+1]\,.
\end{equation}
As to gravity, the theory \eqref{eq:L-modQED-modMax} can be
coupled~\cite{BetschartKK2008,KantKS2009}
to an external gravitational field
(fixed background spacetime metric) but
not to a dynamic gravitational field (variable spacetime metric),
as the energy-momentum tensor is generally
not symmetric~\cite{BetschartKK2008,Kostelecky2004}.

Leaving the gravitational issue aside, return to
the modified-QED theory \eqref{eq:L-modQED-modMax} over a flat
spacetime manifold and ask what parameter values of the single
Lorentz-violating parameter $\widetilde{\kappa}_\text{tr}$
are \emph{a priori} to be expected. It turns out that
simple spacetime-foam models (see Sec.~\ref{sec:Outlook}) can give
positive values of order unity for this deformation parameter,
\begin{equation}
\widetilde{\kappa}_\text{tr}\,\big|^\text{naive\;theory} = \text{O}(1)\,.
\end{equation}
This implies that already the most basic experimental tests of the
effective photon theory \eqref{eq:L-modMax} have the potential
to teach us something of the fundamental properties of
spacetime.

\section{Existing direct laboratory bounds}
\label{sec:Existing-direct-bounds}

The first direct laboratory bound was obtained in
1938 by Ives and Stilwell at Bell Labs, USA,
with the following approximate result~\cite{IvesStilwell1938}:
\begin{equation}
|\widetilde{\kappa}_\text{tr}| \;\lesssim\; 10^{-2}\,.
\end{equation}
Over the years, this difficult experiment has been improved
steadily. The two most recent results were obtained at the
Max-Planck-Institut f\"{u}r Kernphysik
(MPIK) in Heidelberg, Germany~\cite{Reinhardt-etal2007}
and at the University of Western Australia (UWA) in
Perth, Australia~\cite{Hohensee-etal2010},
giving, respectively, the following two--$\sigma$ bounds:
\begin{subequations}
\begin{eqnarray}
|\widetilde{\kappa}_\text{tr}| &<& 2 \times 10^{-7}\,,
\\[2mm]
|\widetilde{\kappa}_\text{tr}| &<& 3 \times 10^{-8}\,.
\end{eqnarray}
\end{subequations}

In principle, this last direct laboratory bound
can be improved by 4 orders of magnitude if cryogenic
resonators are used~\cite{Benmessai2008,Herrmann-etal2009}.
For completeness, also another type of laboratory bound~\cite{Altschul2009}
will be mentioned in the next section.

\section{Indirect bounds from particle astrophysics}
\label{sec:Particle-astrophysics-bounds}

Following up on an early suggestion
by Beall~\cite{Beall1970} $\big($and a later one by
Coleman and Glashow \cite{ColemanGlashow1997}$\big)$, it is possible to
obtain tight \underline{indirect} bounds via particle
astrophysics~\cite{KlinkhamerSchreck2008, Klinkhamer2010}.
The basic idea is remarkably simple~\cite{Beall1970,ColemanGlashow1997}:
\newcounter{alphcounter}     
\begin{list}{(\alph{alphcounter})}{\usecounter{alphcounter}}
\item
With modified dispersion relations,
new decay channels appear which are absent in the standard relativistic theory.
\item
This leads to rapid energy loss of particles with
energies above threshold [a generic LV parameter `$\kappa$'
typically gives a threshold energy $E_\text{thresh}(\kappa)\to\infty$
for $|\kappa|\to 0$\,].
\item
Observing these particles implies that
they necessarily have energies at or below threshold
[$E\leq E_\text{thresh}(\kappa)$], which, in turn, gives bounds
on the LV parameters (`$\kappa$') of the theory.
\end{list}

In modified QED theory
(\ref{eq:L-modQED-modMax}) with $\widetilde{\kappa}_\text{tr} \in (-1,\,1]$,
exact tree-level decay rates have been
calculated for two processes~\cite{KlinkhamerSchreck2008},
which occur for $\widetilde{\kappa}_\text{tr}\ne 0$
because of the difference between the
maximum attainable velocity $c$ of the Dirac particle and
the photon velocity
$c\,\sqrt{(1-\widetilde{\kappa}_\text{tr})/
          (1+\widetilde{\kappa}_\text{tr})}$.
The resulting bounds will be called `indirect,' because they do not
directly rely on the propagation properties of the photon but on
indirect mass-shell effects, as is made clear by point (a)
above.\footnote{Indirect bounds can also be obtained in the
laboratory. For isotropic modified Maxwell theory,
a remarkable bound~\cite{Altschul2009}
has been obtained from the apparent absence
of nonstandard synchrotron-radiation losses
at the Large Electron Positron (LEP) collider of CERN.
This indirect laboratory bound will be
listed in the summary table below.}

The first process, vacuum-Cherenkov radiation
for $\widetilde{\kappa}_\text{tr} > 0\;$ (Fig.~\ref{fig}--left),
is found to have the energy threshold
\begin{subequations}\label{eq:thresh-vacuumCher-photondecay}
\begin{equation}
E_\text{thresh}^\text{(a)} = M\;
\sqrt{\frac{1+\widetilde{\kappa}_\text{tr}}
                 {2\,\widetilde{\kappa}_\text{tr}}}
= \frac{1}{\sqrt{2}}\;\frac{M}{\sqrt{\widetilde{\kappa}_\text{tr}}}
+\mathsf{O}\Big(M\,\sqrt{\widetilde{\kappa}_\text{tr}}\,\Big)\,,
\label{eq:thresh-vacuumCher}
\end{equation}
where the charged particle can be
a proton $p$ or heavy nucleus $N$, each, in first approximation,
considered as a charged pointlike Dirac particle with mass
$M=M_{p}$ or $M=M_{N}$. The vacuum-Cherenkov decay
rate depends, of course, on the
value of the electric charge of the Dirac particle
($|e|$ or $Z_{N}\,|e|$) but the energy threshold does not.
In fact, the energy threshold \eqref{eq:thresh-vacuumCher} simply
follows from energy-momentum conservation.
Still, it is important to know the radiation rate
above threshold, in particular, to make sure that it is not
strongly suppressed.

\begin{figure}[t]
\hspace*{.125\textwidth}
\includegraphics[width=.25\textwidth]{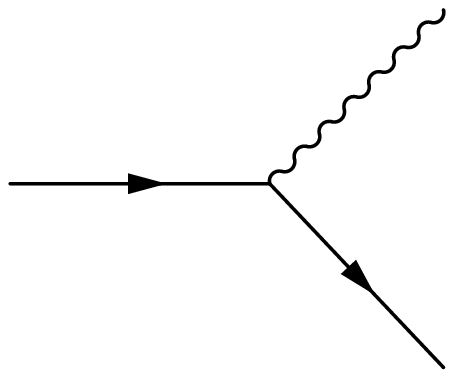}
\hspace*{.25\textwidth}
\includegraphics[width=.25\textwidth]{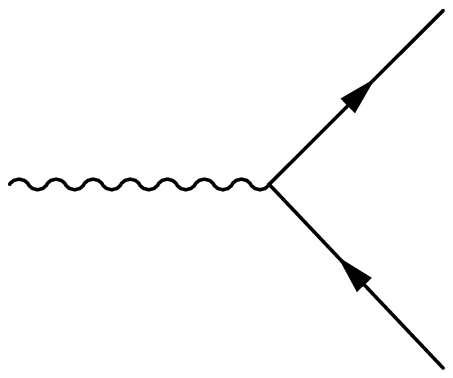}
\hspace*{.0\textwidth}
\vspace*{2mm}
\caption{Left: Vacuum-Cherenkov radiation $p^{+} \to p^{+}\gamma$.
Right: Photon decay $\gamma \to e^{+} e^{-}$.}
\label{fig}
\end{figure}

The second process, photon decay
for $\widetilde{\kappa}_\text{tr} < 0\;$ (Fig.~\ref{fig}--right),
has a similar energy threshold:
\begin{equation}
E_\text{thresh}^\text{(b)}
=2\,M\;\sqrt{\frac{\big(1 -\widetilde{\kappa}_\text{tr}\big)}
               {-2\,\widetilde{\kappa}_\text{tr}}}
=\sqrt{2}\; \frac{M}{\sqrt{-\widetilde{\kappa}_\text{tr}}}
+\mathsf{O}\Big(M\,\sqrt{-\widetilde{\kappa}_\text{tr}}\,\Big),
\label{eq:thresh-photondecay} \end{equation}
\end{subequations}
where the charged particles in the final state can be an
electron and a positron, each considered as a charged pointlike Dirac
particle with mass $M=M_{e}$.

Both decay rates and corresponding energy thresholds are  well-behaved
for parameter values in the domain \eqref{eq:domain}.

\section{Existing indirect earth-based bounds from particle astrophysics}
\label{sec:Existing-indirect-bounds}

The \underline{absence} of vacuum-Cherenkov radiation for a
particular ultra-high-energy-cosmic-ray (UHECR) event~\cite{PAO2010}
from the Pierre Auger Observatory (Auger, for short) with
$E_\text{prim}=(212 \pm 53)\;\text{EeV}$ implies
$E_\text{prim}<E_\text{thresh}^\text{(a)}(\widetilde{\kappa}_\text{tr})$.
Formula (\ref{eq:thresh-vacuumCher}), then, gives the following indirect
two--$\sigma$ upper bound~\cite{KlinkhamerSchreck2008}:%
\begin{subequations}\label{eq:SMEbounds-isotropic}
\begin{equation} \widetilde{\kappa}_\text{tr} \;<\; +0.6 \times 10^{-19}\,,
\label{eq:SMEbounds-isotropic-upper} \end{equation}
for a conservative mass value $M = M_\text{Fe} = 52\;\text{GeV}$.

Similarly, the \underline{absence} of photon decay for
gamma-ray events~\cite{Aharonian-etal2007} from HESS with
$E_{\gamma}=(30\pm 5)\;\text{TeV}$ implies
$E_{\gamma}<E_\text{thresh}^\text{(b)}(\widetilde{\kappa}_\text{tr})$.
Formula (\ref{eq:thresh-photondecay}), then, gives the following
indirect two--$\sigma$ lower bound~\cite{KlinkhamerSchreck2008}:
\begin{equation}
\widetilde{\kappa}_\text{tr} \;>\; - 0.9 \times
10^{-15}\,, \label{eq:SMEbounds-isotropic-lower}
\end{equation}
\end{subequations}
for photon decay into an electron-positron pair with an individual
particle mass $M=M_e = 511\;\text{keV}$.
A similar bound can perhaps be obtained with appropriate
gamma-ray events from VERITAS~\cite{VERITAS}.

Equation (\ref{eq:SMEbounds-isotropic-upper}) or
(\ref{eq:SMEbounds-isotropic-lower})
depends only on the inferred travel length of a meter or more
for the primary at the top of the Earth's atmosphere
and on the energy of this primary. As such, each is an
\underline{earth-based} bound, not an ``astrophysical'' bound.
This type of bound does not depend on the precise (astronomical)
origin of the primary nor on the actual distance between the
source and the Earth. Specifically,
bounds (\ref{eq:SMEbounds-isotropic-upper})--(\ref{eq:SMEbounds-isotropic-lower})
rely on having detected primaries traveling over a few meters in the
Earth's atmosphere and having measured their energy reliably.
(The previous statements on these bounds are somewhat repetitive,
but they hopefully dispel the considerable confusion in the literature
about the nature of these indirect bounds.)

\section{Future indirect earth-based bounds from particle astrophysics}
\label{sec:Future-indirect-bounds}

According to \eqref{eq:thresh-vacuumCher-photondecay},
the bounds from Sec.~\ref{sec:Existing-indirect-bounds}
scale as $(M/E)^2$. This scaling behavior invites the following
considerations.

In the future, it may be possible
to obtain an appropriate Auger sample of UHECR events with%
\begin{subequations}\label{eq:future-samples}
\begin{eqnarray}
E_\text{prim} \;
\stackrel{\boldsymbol{?}}{\boldsymbol{=}} \;
(25\pm 5)\;\text{EeV}\,,\quad M_\text{prim} \;
\stackrel{\boldsymbol{?}}{\boldsymbol{=}}
\; M_p \;=\; 0.938\;\text{GeV}\,.
\end{eqnarray}
Similarly, it may be possible to obtain an appropriate
Cherenkov Telescope Array~\cite{CTA} (CTA)
sample of TeV-gamma-ray events with
\begin{eqnarray}
E_\gamma \;
\stackrel{\boldsymbol{?}}{\boldsymbol{=}} \;
(3.0\pm 0.5)\times 10^2\; \text{TeV}\,.
\end{eqnarray}
\end{subequations}
These data samples would allow us to
improve the previous two--$\sigma$ bounds
(\ref{eq:SMEbounds-isotropic-upper})--(\ref{eq:SMEbounds-isotropic-lower})
by 2 orders of magnitude~\cite{Klinkhamer2010},
\begin{eqnarray}\label{eq:SMEbounds-isotropic-future}
- 0.9 \times 10^{-17}\;
\stackrel{\boldsymbol{?}}{\boldsymbol{<}} \;
\widetilde{\kappa}_\text{tr} \;
\stackrel{\boldsymbol{?}}{\boldsymbol{<}} \;
+ 1.0 \times 10^{-21}\,,
\end{eqnarray}
again with $M=M_e=0.511\;\text{MeV}$ for the lower bound.
The question marks in \eqref{eq:SMEbounds-isotropic-future}
are a reminder that the samples \eqref{eq:future-samples}
are not yet available.

These potential future bounds, together with the existing ones,
are summarized in Table~\ref{tab:bounds-compared}.

\begin{table}[b]
\begin{center}
\renewcommand{\arraystretch}{1.25}   
\begin{tabular}{lcl}
\hline\hline
  Type of bound
& $\widetilde{\kappa}_\text{tr}$
& Experiment + Reference(s)\\
\hline
  Existing, direct
& $\pm 10^{-8}$
& Laboratory: sapphire oscillators, UWA~\cite{Hohensee-etal2010}\\
 Existing, indirect
& $\pm 5 \times 10^{-15}$
& Laboratory: synchrotron 
losses,
LEP (CERN)~\cite{Altschul2009}\\
Existing, indirect & $\big(-10^{-15}\,,\,+10^{-19}\,\big)$
& Particle astrophysics: HESS,\;Auger--S~\cite{KlinkhamerSchreck2008}\\
\hline
Future, direct & $\pm 10^{-12}\,?$
& Laboratory: cryogenic resonators~\cite{Benmessai2008,Herrmann-etal2009}\\
Future, indirect & $\big(-10^{-17}\,?\,,\,+10^{-21}\,?\,\big)$
& Particle astrophysics: CTA,\;Auger--S+N~\cite{Klinkhamer2010}\\
\hline\hline
\end{tabular}
\end{center}
\vspace*{-2mm}
\caption{\mbox{Orders of magnitude for existing and future
two--$\sigma$ bounds on the Lorentz-violating parameter
$\widetilde{\kappa}_\text{tr}$} of isotropic modified Maxwell theory
coupled to standard Dirac theory from laboratory and particle-astro\-physics
experiments (the last experiments only refer to processes
occurring in the Earth's atmosphere).}
\label{tab:bounds-compared}
\end{table}

\section{Outlook}
\label{sec:Outlook}

What do the existing and future bounds on
$\widetilde{\kappa}_\text{tr}$ from Table~\ref{tab:bounds-compared}
imply physically?
Based on very general arguments (Einstein's dynamic spacetime
manifold and Heisenberg's quantum-mechanical uncertainty relations),
Wheeler~\cite{Wheeler1957} has argued that
``quantum spacetime'' must have a nontrivial small-scale structure.
Moreover, it is to be expected that this must
leave \underline{some} remnants (``defects'')
in the effective classical spacetime manifold
relevant over sufficiently large length scales.

Already for rather naive Swiss-cheese-type
classical-spacetime models~\cite{BernadotteKlinkhamer2007},
it has been found that the photon propagation is modified and
corresponds to an isotropic modified Maxwell theory
\eqref{eq:L-modMax} with a positive
deformation parameter of order
\begin{equation} \hspace*{-0mm}
\widetilde{\kappa}_\text{tr}\,\big|^\text{naive\;theory} =
\text{O}\big(\,\widetilde{b}^{\,4}/\,\widetilde{l}^{\:4}\,\big) \geq 0\,,
\label{eq:naive-estimate}
\end{equation}
for a typical defect (``hole'') size $\widetilde{b}$
and a typical separation $\widetilde{l}$
between the individual defects (holes) randomly embedded
in Minkowski spacetime. Equally important, the same defects
do \underline{not} modify the maximum velocity of the Dirac
particle, at least to leading order~\cite{BernadotteKlinkhamer2007}.
This implies that the modified QED theory \eqref{eq:L-modQED-modMax}
corresponds to the \underline{effective theory} of standard photons
and Dirac particles propagating over a Swiss-cheese-type
classical spacetime. In principle, it may be that the
defects (holes) have sizes and separations related by
$\widetilde{b} \lesssim \widetilde{l} \lesssim
l_\text{\,Planck}  \equiv \sqrt{\hbar\,G/c^3}
\approx 1.6 \times 10^{-35}\,\text{m}$
or by $\widetilde{b} \lesssim \widetilde{l} \lesssim l$
if ``quantum spacetime'' has a new fundamental length
$l$ as argued in Ref.~\cite{Klinkhamer2007}.

 From \eqref{eq:naive-estimate},
the suggestion is that the physically relevant quantity is
perhaps not $\widetilde{\kappa}_\text{tr}\geq 0$
but rather its quartic root,
\vspace*{-1mm}
\begin{equation}
\big(\,\widetilde{\kappa}_\text{tr}\,\big)^{1/4}\geq 0\,.
\end{equation}
Taking values at the boundaries of the range of
Table~\ref{tab:bounds-compared}, observe that, on the one hand,
the number $(10^{-8})^{1/4}=10^{-2}$ is small
but not very small and, on the other hand,
the number $(10^{-20})^{1/4}=10^{-5}$ really is very small
[of the same order as the ratio of the nucleus radius
over the atomic radius].

Particle astrophysics thus provides a null experiment
suggesting that spacetime is \underline{unexpectedly}
\underline{smooth}
[quantified as $\widetilde{b}/\widetilde{l} \lesssim 10^{-5}$
for the effective parameters $\widetilde{b}$ and $\widetilde{l}$
mentioned below \eqref{eq:naive-estimate}].
Perhaps this null experiment from particle astrophysics will turn out
to be as important as the Michelson--Morley
experiment~\cite{MichelsonMorley1887},
which led to Einstein's radically new concept of
the relativity of simultaneity
and the special theory of relativity~\cite{Einstein1905}.
Also in our case, it appears that radically new concepts are needed
to understand the nature of what we call, for convenience,
``quantum spacetime'' but which may have an entirely novel content.

\acknowledgments

The author thanks the organizers of the conference and
those of the particle-astrophysics session for their
kind hospitality. 
He also thanks his many collaborators of the last years.

\newcommand{\textjournal}{\text} 


\begin{thebibliography}{99}

\bibitem{ChadhaNielsen1983}
S. Chadha  and  H.B.~Nielsen,
\textit{Lorentz invariance as a low-energy phenomenon},
\textjournal{Nucl. Phys. B} \textbf{217}  (1983) 125.

\bibitem{KosteleckyMewes2002}
V.A.~Kosteleck\'{y}  and  M.~Mewes,
\textit{Signals for Lorentz violation in electrodynamics},
\textjournal{Phys. Rev. D} \textbf{66}  (2002) 056005
[\texttt{arXiv:hep-ph/0205211}].

\bibitem{KlinkhamerSchreck2010}
F.R.~Klinkhamer  and  M.~Schreck,
\textit{Consistency of isotropic modified Maxwell theory: Microcausality and
unitarity},
to appear in \textjournal{Nucl. Phys. B}
[\texttt{arXiv:1011.4258}].  

\bibitem{BetschartKK2008}
G.~Betschart, E.~Kant, and F.R.~Klinkhamer,
\textit{Lorentz violation and black-hole thermodynamics},
\textjournal{Nucl. Phys.  B} \textbf{815} (2009) 198
[\texttt{arXiv:0811.0943}]. 

\bibitem{KantKS2009}
E.~Kant, F.R.~Klinkhamer, and M.~Schreck,
\textit{Lorentz violation and black-hole
  thermodynamics: Compton scattering process},
\textjournal{Phys. Lett.  B} \textbf{682} (2009) 316
[\texttt{arXiv:0909.0160}]. 

\bibitem{Kostelecky2004}
V.A. Kosteleck\'{y},
\textit{Gravity, Lorentz violation, and the standard model},
\textjournal{Phys. Rev.  D} \textbf{69} (2004) 105009
[\texttt{arXiv:hep-th/0312310}].

\bibitem{IvesStilwell1938}
 H.E. Ives  and  G.R. Stilwell,
\textit{An experimental study of the rate of a moving atomic clock},
\textjournal{J. Opt. Soc. Am.} \textbf{28}  (1938) 215.

\bibitem{Reinhardt-etal2007}
S. Reinhardt  et al.,
\textit{Test of relativistic time dilation with fast optical atomic
  clocks at different velocities},
\textjournal{Nature Phys.} \textbf{3}  (2007) 861.

\bibitem{Hohensee-etal2010}
M.A.~Hohensee  et al.,
\textit{Improved constraints on isotropic shift and anisotropies of the speed of
  light using rotating cryogenic sapphire oscillators},
\textjournal{Phys. Rev. D}  \textbf{82} (2010) 076001
[\texttt{arXiv:1006.1376}].  

\bibitem{Benmessai2008}
K.~Benmessai et al.,
\textit{Measurement of the fundamental thermal noise limit in a cryogenic sapphire
  frequency standard using bimodal maser oscillations},
\textjournal{Phys. Rev. Lett.} \textbf{100} (2008) 233901.


\bibitem{Herrmann-etal2009}
S.~Herrmann, A.~Senger, K.~M\"{o}hle, M.~Nagel, E.V.~Kovalchuk, and A.~Peters,
\textit{Rotating optical cavity experiment testing Lorentz invariance at the
  $10^{-17}$ level},
\textjournal{Phys. Rev. D} \textbf{80} (2009) 105011
[\texttt{arXiv:1002.1284}].  

\bibitem{Altschul2009}
B.~Altschul,
\textit{Bounding isotropic Lorentz violation using synchrotron losses
 at LEP},
\textjournal{Phys. Rev.  D} \textbf{80} (2009) 091901
[\texttt{arXiv:0905.4346}]. 

\bibitem{Beall1970}
E.F.~Beall,
\textit{Measuring the gravitational interaction of elementary particles},
\textjournal{Phys.\,Rev.\,D}\,\textbf{1}\,(1970)\,961. 

\bibitem{ColemanGlashow1997}
S.R.~Coleman  and  S.L.~Glashow,
\textit{Cosmic ray and neutrino tests of special relativity},
\textjournal{Phys. Lett. B} \textbf{405} (1997) 249
[\texttt{arXiv:hep-ph/9703240}].


\bibitem{KlinkhamerSchreck2008}
F.R.~Klinkhamer  and  M.~Schreck,
\textit{New two-sided bound on the isotropic Lorentz-violating parameter of modified Maxwell theory},
\textjournal{Phys. Rev. D}  \textbf{78} (2008) 085026
[\texttt{arXiv:0809.3217}].

\bibitem{Klinkhamer2010}
F.R.~Klinkhamer,
\textit{Potential sensitivities to Lorentz violation from nonbirefringent modified
  Maxwell theory of Auger, HESS, and CTA},
\textjournal{Phys. Rev. D} \textbf{82} (2010) 105024
[\texttt{arXiv:1008.1967}]. 

\bibitem{PAO2010}
J.~Abraham  et al. [Pierre Auger Observatory Collaboration],
\textit{Measurement of the depth of maximum of extensive air showers above $10^{18}$ eV},
\textjournal{Phys. Rev. Lett.} \textbf{104} (2010) 091101
[\texttt{arXiv:1002.0699}];
for further information on the observatory,
see \texttt{http://www.auger.org/}.

\bibitem{Aharonian-etal2007}
F.~Aharonian  et al. [HESS Collaboration],
\textit{Primary particle acceleration above 100 TeV in the shell-type supernova
  remnant RX J1713.7--3946 with deep H.E.S.S. observations},
\textjournal{Astron. Astrophys.}  \textbf{464} (2007) 235
[\texttt{arXiv:astro-ph/0611813}];
for further information on the telescope array,
see \texttt{http://www.mpi-hd.mpg.de/hfm/HESS/}.

\bibitem{VERITAS}
T.C.~Weekes et al. [VERITAS Collaboration],
\textit{VERITAS: the Very Energetic Radiation Imaging
        Telescope Array System},
\textjournal{Astropart. Phys.} {\bf 17}, 221 (2002)
[\texttt{arXiv:astro-ph/0108478}];\newline
see also \texttt{http://veritas.sao.arizona.edu/}.


\bibitem{CTA}
G.~Hermann, W.~Hofmann, T.~Schweizer, and M.~Teshima  [CTA Consortium],
\textit{Cherenkov Telescope Array: The next-generation ground-based gamma-ray
  observatory},
\textjournal{Proc. 30th Int. Cosm. Ray Conf.}
\textjournal{(ICRC 2007)}  
[\texttt{arXiv:0709.2048}];
see also \texttt{http://www.cta-observatory.org/}.

\bibitem{Wheeler1957}
J.A. Wheeler,
\textit{On the nature of quantum geometrodynamics},
\textjournal{Ann. Phys. (N.Y.)}  {\bf 2}, 604 (1957).

\bibitem{BernadotteKlinkhamer2007}
S. Bernadotte  and  F.R. Klinkhamer,
\textit{Bounds on length scales of classical spacetime foam models},
\textjournal{Phys. Rev. D} \textbf{75} (2007) 024028
[\texttt{arXiv:hep-ph/0610216}].

\bibitem{Klinkhamer2007}
F.R.~Klinkhamer,
\textit{Fundamental length scale of quantum spacetime foam},
\textjournal{JETP Lett.}  {\bf 86} (2007)  73
[\texttt{arXiv:gr-qc/0703009}].

\bibitem{MichelsonMorley1887}
A.A. Michelson and E.W. Morley,
\textit{On the relative motion of the Earth and the luminiferous ether},
\textjournal{Am. J. Sci.}   {\bf 34}  (1887) 333.

\bibitem{Einstein1905}
A. Einstein,
\textit{Zur Elektrodynamik bewegter Körper},
\textjournal{Ann. Phys. (Leipzig)} {\bf 17}, 891 (1905)
[reprinted in: \textjournal{Ann. Phys. (Leipzig)} {\bf 14}, S1, 194 (2005)].

\end{thebibliography}
\end{document}